**PAPER**

Check for updates

# High polarization, endurance and retention in sub-5 nm Hf$_{0.5}$Zr$_{0.5}$O$_2$ films†

Jike Lyu, Tingfeng Song, Ignasi Fina [ID] * and Florencio Sánchez [ID] *

Ferroelectric HfO$_2$ is a promising material for new memory devices, but significant improvement of its important properties is necessary for practical application. However, previous literature shows that a dilemma exists between polarization, endurance and retention. Since all these properties should be simultaneously high, overcoming this issue is of the highest relevance. Here, we demonstrate that high crystalline quality sub-5 nm Hf$_{0.5}$Zr$_{0.5}$O$_2$ capacitors, integrated epitaxially with Si(001), present combined high polarization (2$P_r$ of 27 μC cm$^{-2}$ in the pristine state), endurance (2$P_r$ > 6 μC cm$^{-2}$ after 10$^{11}$ cycles) and retention (2$P_r$ > 12 μC cm$^{-2}$ extrapolated at 10 years) using the same poling conditions (2.5 V). This achievement is demonstrated in films thinner than 5 nm, thus opening bright possibilities in ferroelectric tunnel junctions and other devices.

## Introduction

One of the greatest virtues of the recently discovered[1] ferroelectricity in orthorhombic HfO$_2$ is the robustness of the ferroelectricity in nanometric layers.[2] Usually, there is a peak in the dependence of polarization on film thickness. The largest polarization is observed in films ~8–10 nm thick.[3–6] High remnant polarization up to ~30 μC cm$^{-2}$ has been reported in polycrystalline HfO$_2$-based films of thickness ($t$) = 10–16 nm (ref. 7–10) and epitaxial $t$ = 9 nm (ref. 11) and 5 nm (ref. 12) Hf$_{0.5}$Zr$_{0.5}$O$_2$ (HZO) films. Films thinner than 5 nm have been reported to be ferroelectric,[3,13–15] which can permit their use in ferroelectric tunnel junctions,[17–19] but simultaneous (same sample and same poling voltage) combined high polarization, endurance and retention have not been reported for sub-5 nm films.

The lower endurance of ferroelectric HfO$_2$ compared to conventional ferroelectric perovskites[20–22] is a limitation. HfO$_2$ capacitors can be optimized to enhance the endurance, but it occurs at the cost of reduction of polarization. This is called the polarization-endurance dilemma.[2] Significant enhancement by doping HZO with 1 mol% La was reported by Chernikova et al.,[10] with a maximum remnant polarization of 15 μC cm$^{-2}$ and maintaining a $P_r$ higher than 12 μC cm$^{-2}$ for up to 4 × 10$^{10}$ cycles without breakdown in $t$ = 10 nm films. More recently, Kozodaev et al.[9] reported an endurance of 10$^{11}$ cycles for 0.7 mol% La doped HZO films, 10 nm thick, without degradation of polarization and a very large $P_r$ of 14 μC cm$^{-2}$ after 10$^{11}$ cycles. Later, Mehmood et al.[23] revealed the presence of the endurance-retention dilemma in La doped films, reporting that the optimization of endurance by La content causes retention loss. Enhancement of endurance is also achieved by using a low cycling voltage, but it causes reduction of remnant polarization and, consequently, retention.[11,18,21] The case of ultrathin films (<5 nm) is particularly dramatic. In ultrathin capacitors, depolarization fields and current leakage (particularly at pinholes and grain boundaries) have huge detrimental influence and long retention above 10 years could be unattainable. The combination of high polarization, endurance and retention is necessary, but surprisingly, the study of these properties is usually addressed separately. In particular, retention and endurance under same operation voltage are scarcely compared and, to the best of our knowledge, have not been reported for films thinner than 5 nm. Thus, it is of major interest to achieve combined large polarization-endurance-retention in ultrathin ferroelectric films based in HfO$_2$.

*Institut de Ciència de Materials de Barcelona (ICMAB-CSIC), Campus UAB, Bellaterra 08193, Barcelona, Spain. E-mail: ifina@icmab.es, fsanchez@icmab.es*

†Electronic supplementary information (ESI) available: XRD scans and AFM images, current leakage curves, polarization loops measured by PUND. Imprint voltage data. Retention measurements and fitting. Comparison of the endurance of all the films. Experimental *J–V* loops and simulation. Extended information on endurance and retention. See DOI: 10.1039/d0nr02204g

Epitaxial HZO films, with well-controlled microstructure and high polarization,[11,13,17–19,24–27] are appropriate to achieve insight into the limits of the polarization-endurance-retention dilemma in ultra-thin ferroelectric HfO$_2$. Here, we report on the ferroelectric polarization, endurance, and retention of a series of epitaxial HZO films on Si(001) with thickness in the $t$ = 4.6–18.4 nm range. The films are ferroelectric with monotonic increase in polarization up to above 30 μC cm$^{-2}$ as thickness is reduced. A film thinner than 5 nm operating at 2.5 V has a 2$P_r$ of 27 μC cm$^{-2}$, endurance up to at least 10$^{11}$ cycles without breakdown (although fatigue reduces $2P_r$ to 6.6 μC cm$^{-2}$) and retention extending well beyond 10 years. The coexistence of high polarization, endurance and retention under the same operating voltage in films thinner than 5 nm demonstrates the promising potential of ferroelectric HfO$_2$ for new memories, including tunneling devices.

## Results and discussion

HZO films were integrated epitaxially on Si(001) using the oxide heterostructure La$_{2/3}$Sr$_{1/3}$MnO$_3$ (LSMO)/LaNiO$_3$ (LNO)/ CeO$_2$/yttria-stabilized zirconia (YSZ) as a buffer layer (Fig. 1a).[24] The first layer, YSZ, was used as an epitaxial template[28] and the top conducting oxide LSMO layer acts as a bottom electrode. The other layers are used to accommodate the lattice mismatch. The X-ray diffraction (XRD) $\theta$–$2\theta$ scan of the thinnest HZO film ($t$ = 4.6 nm) is shown in Fig. 1b. There are high intensity (00 l) reflections of the Si substrate and the YSZ, CeO$_2$, LNO and LSMO buffer layers. The (111) reflection of orthorhombic HZO is observed at $2\theta \approx 30°$. The scan is similar to that of a thicker HZO film on LSMO/LNO/CeO$_2$/YSZ/ Si(001),[24] although in Fig. 1b the o-HZO(111) peak has lower intensity because its thickness is less. The $\theta$–$2\theta$ scans of the other samples in the series are similar (Fig. S1, ESI†), showing the intensity of the o-HZO(111) peak increasing with thickness. Fig. 1c shows $2\theta$–$\chi$ frames of the samples of HZO thicker than around 7 nm. The o-HZO(111) reflection in the $2\theta$–$\chi$ frames is a bright circular spot in all samples, according to its epitaxial growth.[24] The corresponding $\chi$-scans (Fig. S2, ESI†) are narrow peaks, with full-width at half-maximum (FWHM) around 2°. The FWHM value increases only very slightly with thickness, suggesting that the crystallinity of the o-HZO phase is preserved in the range of film thicknesses investigated. The XRD pole figure from the o-HZO(−111) and Si(111) reflections of the $t$ = 9.7 nm sample was also measured (Fig. S2, ESI†). The pole figure confirms epitaxy, with the presence of four in-plane HZO crystal variants and a [1−10]HZO(111)//[100]Si(001) epitaxial relationship, as reported for epitaxial HZO films on SrTiO$_3$ (ref. 11) and YSZ[24] buffered Si(001). Monoclinic reflec-tions in the $2\theta$–$\chi$ frames are below the detection limit in the $t$ = 6.9 nm film, but (−111) and (002) spots, elongated along $\chi$, are observed in the thicker films, with a strong increase in intensity with thickness. Additionally, topographic atomic force microscopy (AFM) images (Fig. S3, ESI†) show flat surfaces in all films, with very low root-mean-square (rms) roughness in the 0.2–0.5 nm range.

All HZO films show ferroelectric polarization hysteresis loops (Fig. 2a). Remnant polarization decreases monotonically with thickness, from about 33 μC cm$^{-2}$ ($t$ = 4.6 nm film) to 10 ($t$ = 18.4 nm film). The maximum electric field to measure the loops was limited by the breakdown field, which is smaller in the thicker films in spite of the reduction of current leakage with thickness (Fig. S4, ESI†). It is noted that electrical breakdown is initiated by increased transport through specific conduction channels. Similar diminution of electrical breakdown with thickness is usually observed in ferroelectric oxides.[29] Polarization loops of selected samples measured by applying various maximum voltages are shown in Fig. S5, ESI.† The loops of the thinner films show a rounded shape at the largest applied voltages, signaling extrinsic contributions. Some of the thinnest films were also measured by the positive-up negative-down (PUND) method (Fig. S6, ESI†). The remnant polarizations of the $t$ = 4.6, 6.9 and 9.7 nm films as measured by the PUND method are slightly lower, around 2, 5 and 0.1 μC cm$^{-2}$, respectively. Note that the polarizations measured by PUND can be underestimated due to saturated loops not being measured because of the observed increase of coercive field when using PUND technique (Fig. S6, ESI†) due to the fluid imprint field,[30] probably caused by charge injection. On the other hand, Fig. 2a shows that the loops are shifted to negative voltages, which corresponds to an imprint field pointing from the top Pt electrode towards the bottom LSMO electrode (inset of Fig. 2a). The imprint voltage is −0.08 V in the $t$ = 4.6 nm film and increases in magnitude with film thickness up to around −0.36 V in the $t$ = 6.9 nm film; it is similar in the thicker films (Fig. S7, ESI†). The dielectric permittivity depends largely on the HZO thickness (Fig. 2b). The $t$ = 4.6 nm film shows a butterfly shaped hysteresis loop with permittivity values in the 33.5–37.5 range. Permittivity hysteresis decreases

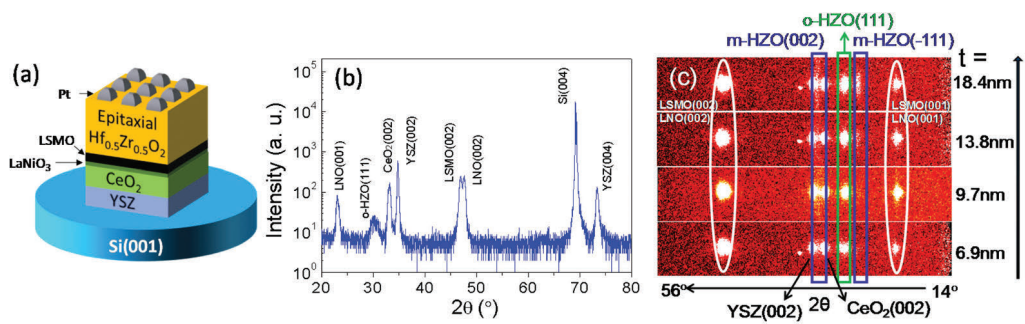

**Fig. 1** (a) Sketch of the studied structure. (b) XRD $\theta$–$2\theta$ symmetric scan of the 4.6 nm HZO film. (c) XRD $2\theta$–$\chi$ frames of samples of selected HZO thicknesses (indicated at the right). The reflections of orthorhombic and monoclinic HZO are marked with green and blue boxes, respectively. The bright spot at $2\theta \sim 36.5°$ present in some frames is a spurious signal caused by the 2D detector.

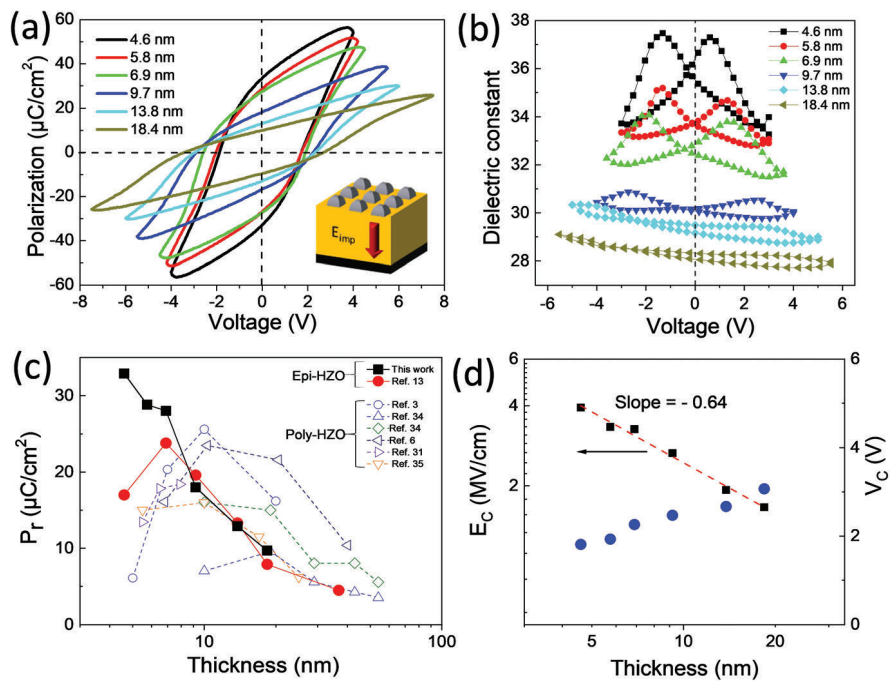

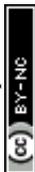

**Fig. 2** (a) Ferroelectric polarization and (b) dielectric permittivity loops. A sketch of the imprint ferroelectric field direction is inset in (a). (c) Remnant polarization plotted as a function of the HZO thickness of the samples reported in this work (black solid squares). The figure shows also data (red solid circles) of epitaxial films grown using the same conditions on LSMO/SrTiO₃(001).[13] Empty symbols correspond to several thickness series reported in literature for polycrystalline HZO films.[3,6,13,31,34,35] References are indicated in the top right of the figure. Most of the reported $P_r$ values were estimated from graphs in the cited references. (d) Coercive electric field (black squares, left vertical axis in logarithmic scale) and coercive voltage (blue circles, right vertical axis) plotted as a function of the HZO thickness (logarithmic scale).

progressively with thickness, although the butterfly shape is clear up to a thickness of around 10 nm. Permittivity values also decrease with thickness, with the thickest film ($t = 18.4$ nm) having permittivity around 28 and almost no hysteresis. The butterfly loops are shifted towards negative values as are the polarization loops, with a similar imprint field. The permittivity of the thinnest films, 32–34 at saturation, is similar to that reported in orthorhombic hafnia.[31–33] The decrease in polarization and permittivity with thickness is possibly due to the greater amount of monoclinic phase, which has null remnant polarization and lower permittivity.[33] The dependence on thickness is, therefore, in agreement with the observation (Fig. 1c) of a larger fraction of monoclinic phase in thicker films.

The remnant polarization is plotted as a function of thickness in Fig. 2c (solid black squares). The monotonic increase with decreasing HZO thickness contrasts with the recurrent observation of reduced polarization in thinner hafnia films, *i.e.* a peaky dependence.[3–6] Data from several thickness series reported in literature[3,6,13,31,34,35] for polycrystalline HZO are shown in the graph (empty symbols). Polycrystalline films present smaller polarization, particularly when they are thinner than around 10 nm. Fig. 2c also depicts the remnant polarization of epitaxial HZO films on SrTiO₃(001) deposited using same conditions as the epitaxial films on Si(001).[13] Thus, Fig. 2c shows that polarization in ultrathin epitaxial

HZO films on Si(001) is largely enhanced with respect to polycrystalline films and epitaxial films on SrTiO₃(001).

Fig. 2d shows the dependence of $E_c$ on thickness. The coercive field could be conditioned by the amount of monoclinic phase, which increases with thickness. However, the recent demonstration[27] of control of the ratio between monoclinic and orthorhombic phases by epitaxial stress (*via* substrate selection) discards this possibility. In ref. 27, films of same thickness deposited on substrates with different lattice parameters presented very different amounts of monoclinic phase and yet the coercive field was very similar. It is also observed that the here reported epitaxial HZO films on Si(001) display $E_c \sim t^{-2/3}$ scaling dependence. In polycrystalline hafnia films, the coercive electrical field does not vary significantly with thickness,[6] whereas $E_c \sim t^{-2/3}$ scaling is usual in high quality perovskite films.[28,29,36,37] The $E_c \sim t^{-2/3}$ scaling is also observed in epitaxial HZO films grown on SrTiO₃(001),[13] pointing to an intrinsic origin in high quality hafnia films.

The retention of polarization (Fig. 3a) was measured for positive (solid symbols) and negative (empty symbols) poling voltages. Each sample was poled, applying the maximum voltage used for measuring the ferroelectric polarization loops (Fig. S8, ESI†), and the polarization was read after the indicated delay time ($t_d$). The retention of some of the films was also measured for different poling voltages and longer times (Fig. S9, ESI†). Retention notably depends on thickness and is

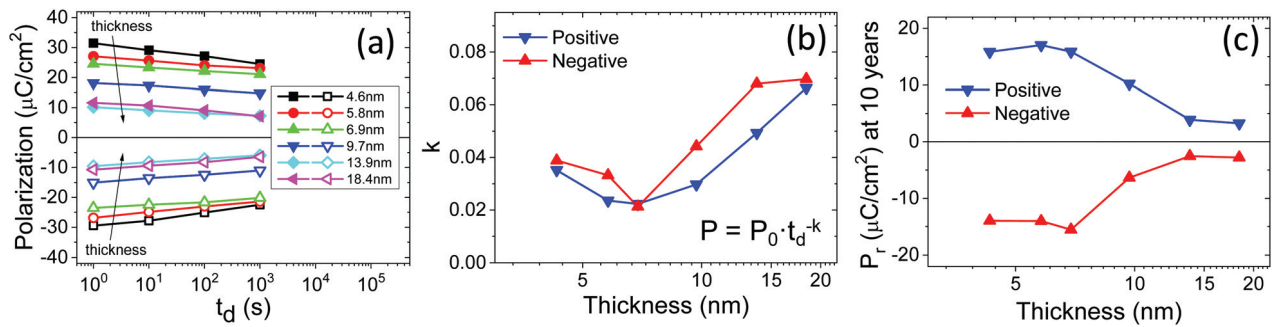

**Fig. 3** (a) Polarization retention after positive (solid symbols) and negative (empty symbols) poling. (b) Exponent $k$, for positive and negative poling, calculated from the retention graphs plotted as a function of HZO thickness. (c) Extrapolated remnant polarization at 10 years using the data fitted to the $P_r = P_0 t_d^{-k}$ equation (Fig. S10, ESI†).

very long in films thinner than 10 nm. The observed large retention in ultrathin films is a consequence of the large $E_c$, characteristic of ferroelectric hafnium films.[38,39] $P_r$ dependence on time has been fitted to the rational dependence $P_r = P_0\ t_d^{-k}$ (Fig. S10, ESI†),[40,41] where $t_d$ refers to the time after poling. The $k$ dependence on thickness and poling direction is shown in Fig. 3b. It is seen that retention in all films is longer (smaller $k$) for positive poling (downwards polarization) than for negative poling (upwards polarization). This is likely due to the discussed imprint field, which makes downward polarization more stable. The retention is longer ($k$ is smaller) with decreasing thickness. This can be caused by the decreased fraction of the monoclinic phase. The coexistence of paraelectric monoclinic and ferroelectric orthorhombic phases can create depolarizing fields in the bulk film if the local polarization vector has an in-plane component.[23] This is the case in the HZO(111) epitaxial films and thus the presence of the monoclinic phase could induce a depolarizing field. In addition, grain boundaries between monoclinic and orthorhombic phases (more abundant in thicker films) can act as conduction pathways for slow ionic transport, with severe impact for retention under a depolarizing field. The exponent $k$ is minimal (longest retention) for the $t = 6.9$ nm film. Further thickness reduction results in degradation of the retention ($k$ is higher), most probably due to the larger leakage current (Fig. S4, ESI†). The extrapolated $P_r$ at 10 years for all films using the $P_r = P_0\ t_d^{-k}$ dependence is shown in Fig. 3c. The thinnest films (up to 6.9 nm) present extrapolated remnant polarization of ≈15 μC cm⁻², which decreases to less than 4 μC cm⁻² with increasing thickness.

Fig. 4a–c shows the endurance characterized at 100 kHz. Fatigue is severe in the thickest film, $t = 18.4$ nm, with quick

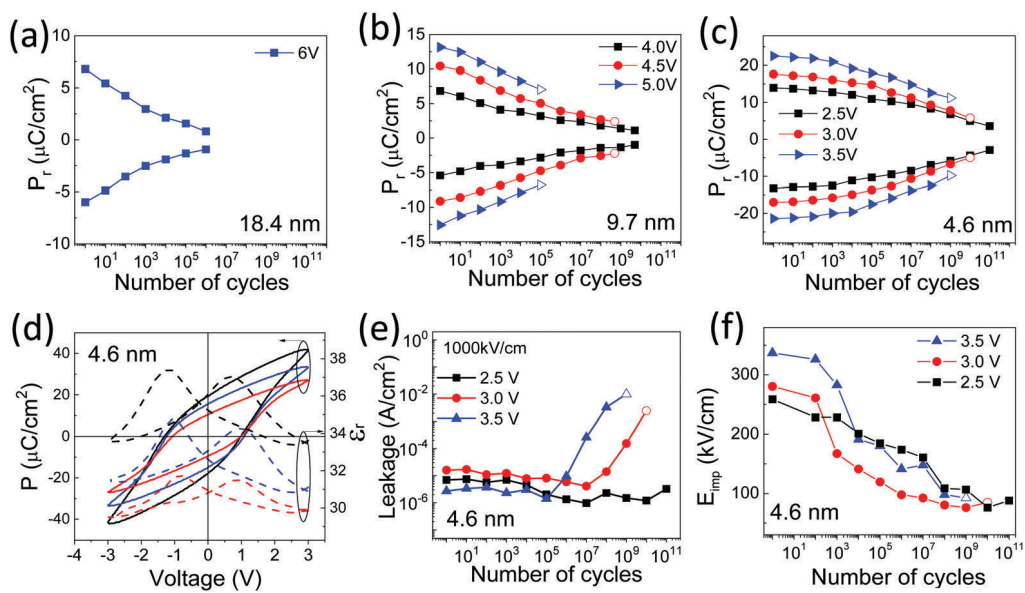

**Fig. 4** Endurance of the $t = 18.4$ nm (a), 9.7 nm (b), and 4.6 nm (c) films. Poling voltages are indicated in the labels of the corresponding graphs. Empty symbols signal hard breakdown of the capacitors. (d) Polarization and permittivity loops of the $t = 4.6$ nm HZO film in pristine condition (black lines) and after $10^4$ (blue lines) and $10^8$ (red lines) cycles at 3 V. Current leakage (e) and imprint field (f) of the $t = 4.6$ nm HZO film as a function of the number of cycles at amplitudes 2.5 V (black squares), 3 V (red circles) and 3.5 V (blue triangles).

degradation of the remnant polarization during switching (Fig. 4a) and a $2P_r$ less than 2 μC cm$^{-2}$ after $10^6$ cycles at 6 V (~3.3 MV cm$^{-1}$). Thinner films are much more robust during cycling. The $t = 9.7$ nm film can be switched for up to $5 \times 10^9$ cycles at 4 V (~4.1 MV cm$^{-1}$) before $2P_r$ is reduced to 2 μC cm$^{-2}$ (Fig. 4b). The initial polarization is larger when increasing the switching voltage, but the capacitors suffer hard breakdown (empty symbols) after a lower number of cycles ($5 \times 10^8$ at 4.5 V and $1 \times 10^5$ at 5 V). Endurance improves with further decrease in HZO thickness and the thinnest film, $t = 4.6$ nm, exhibits excellent behavior against fatigue (Fig. 4c). When it is cycled at 3.5 V (~7.6 MV cm$^{-1}$), the high initial $2P_r$ polarization around 45 μC cm$^{-2}$ diminishes progressively and the capacitor displays a $2P_r > 20$ μC cm$^{-2}$ before hard breakdown after $10^9$ cycles. At 3.0 V (~6.5 MV cm$^{-1}$), hard breakdown occurs after $10^{10}$ cycles. The film can be switched more times at lower voltages and, after $1 \times 10^{11}$ cycles at 2.5 V (~5.4 MV cm$^{-1}$), the $2P_r$ is >6 μC cm$^{-2}$ and the capacitor does not undergo breakdown. The plot of the normalized remnant polarization of all the samples and after $10^4$ cycles is summarizes the impact of thickness on ferroelectric fatigue (Fig. S11, ESI†). Note that the films with $t > 4.6$ nm show no wake-up effect. Instead, for the thinnest sample, $t = 4.6$ nm, wake-up effect is visible in the current–voltage curves but does not affect the polarization value (Fig. S12, ESI†). Absence of wake-up is common in epitaxial films grown on different substrates,[11,24,26] in contrast with polycrystalline hafnia films. The different behaviors can be due to either the different microstructure (epitaxial or polycrystalline films) or the different electrodes (LSMO or TiN bottom electrodes). TiN electrodes, commonly used in polycrystalline films, can induce formation of oxygen vacancies in their proximity,[42] whereas LSMO is not expected to severely reduce the hafnia film.

Polarization loops of the $t = 4.6$ nm HZO film in pristine state and after $10^4$ and $10^8$ cycles at 3 V are presented in Fig. 4d. In spite of the fatigue, the polarization loop after $10^8$ cycles is well saturated and $2P_r$ remains high (~18 μC cm$^{-2}$). The rounder shape near the maximum applied field of the less-cycled loops finds its origin in the broader switching occurring at these stages, which decreases with cycling, reminiscence of the wake-up effect (Fig. S12, ESI†). Also, decreased leakage at initial sample cycling (discussed below) and other extrinsic effects[43] might contribute (Fig. S12, ESI†) to the round shape. The permittivity loops also preserve the butterfly shape, but with smaller values ($\varepsilon \approx 32.5$ at 3 V in the pristine state, diminishing to ~30 after $10^8$ cycles). Similar reduction of permittivity has been reported for fatigued polycrystalline doped-HfO$_2$.[32,42,44]

As mentioned, the $t = 4.6$ nm capacitors suffered hard breakdown after $10^9$ and $10^{10}$ cycles at 3.5 and 3 V, respectively (Fig. 4c). The evolution of the current leakage during cycling is presented in Fig. 4e. During the first cycles, there is a small reduction in leakage current. Afterwards, there is an abrupt increase in leakage after around $10^5$ and $10^7$ cycles at 3.5 and 3 V, respectively. The capacitors cycled at 2.5 V preserve low leakage after $10^{11}$ cycles without breakdown. The imprint field decreases progressively upon cycling (Fig. 4f) from around 300 kV cm$^{-1}$ in the pristine state to around 100 kV cm$^{-1}$ for the three poling voltages. Similar reduction with cycling was reported for polycrystalline ferroelectric HZO capacitors.[45] In brief, there is not a close relation between the sudden leakage increase and the progressive polarization/imprint reduction under cycling. Charged defect (oxygen vacancy) redistribution under cycling is the most plausible scenario to explain the observed progressive polarization[42] and imprint reduction. In perovskite ferroelectric films, it is reported that oxygen vacancies can pin ferroelectric domains.[29] This is also in agreement with the small reduction of leakage during the first cycles, due to mobile charges being pinned. The same charged defects are at the origin of the leakage increase, but in this case generating conduction pathways and finally producing the dielectric breakdown.

The large endurance (Fig. 4c) of the sub-5 nm film ($10^{11}$ cycles) is obtained at an electric field of around 5.5 MV cm$^{-1}$, showing remnant polarization close to 14 μC cm$^{-2}$ in the pristine state. Also, good endurance results are the combination of 17 μC cm$^{-2}$–$10^{10}$ cycles and 22 μC cm$^{-2}$–$10^9$ cycles at 6.5 and 7.6 MV cm$^{-1}$, respectively, obtained for the same sample. To better understand the meaning of these values obtained in a sub-5 nm film, Fig. 5 summarizes the polarization and endurance values reported in literature[6,8–10,46–59] for polycrystalline HfO$_2$ films doped with different atoms. The recent use of La as a dopant has allowed a notable increase in endurance without loss of polarization (empty triangles) in >5 nm films.[9,10] The combined large polarization and endurance in the <5 nm epi-taxial HZO films (solid red stars) are comparable to those shown by La doped polycrystalline HZO films. However, detailed analysis shows that fatigue and polarization in epitaxial films degrade from $P_r \sim 13.3$ to $\sim 3.3$ $\mu$C cm$^{-2}$ after $10^{11}$ cycles, whereas La doped polycrystalline films (>5 nm)[9] show an initial $P_r$ of $\sim 8$ $\mu$C cm$^{-2}$ that increases to $\sim 14$ $\mu$C cm$^{-2}$ in the awake state after $\sim 10^7$ cycles and is without fatigue after $10^{11}$ cycles, thus showing a larger $P_r$ than in epitaxial HZO films after a large cycling number. Nevertheless, the excellent properties of epitaxial HZO films are achieved in films thinner than 5 nm and, despite the ultrathin thickness, using the same poling voltage of 2.5 V, the polarization $2P_r > 12$ $\mu$C cm$^{-2}$ is retained after 10 years (Fig. S9, ESI†).

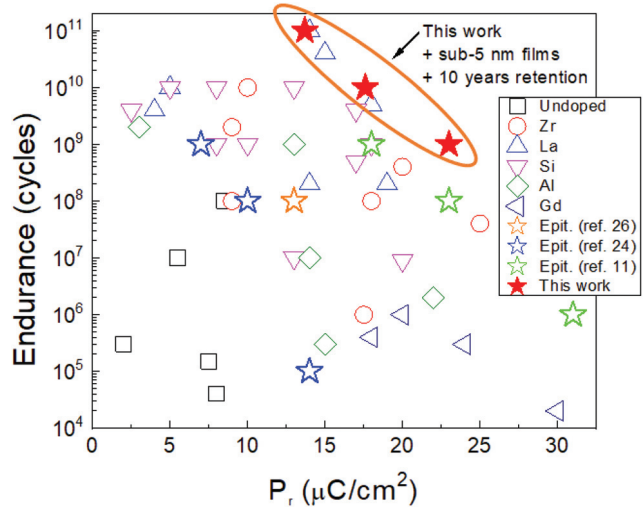

**Fig. 5** Endurance and remnant polarization reported in literature for HfO$_2$ films undoped or with various dopants. The remnant polarization is the maximum value (after wake up in the case of polycrystalline films). Stars correspond to epitaxial films.[11,24,26] Solid red stars correspond to the $t = 4.6$ nm epitaxial film reported in this work. Other symbols correspond to various polycrystalline films: undoped HfO$_2$,[46–48] Zr-doped,[6,16,31,49] La-doped HfO$_2$ and HZO,[8–10,50] Si-doped,[8,51–55] Al-doped,[8,56,57] and Gd-doped.[58,59] Some of the reported endurance and $P_r$ values were estimated from graphs of the cited references.

We note that research on doped HfO$_2$ epitaxial films is in elementary stages in comparison with polycrystalline films. The critical roles of deposition parameters[13] and epitaxial stress[27] on the stabilization of the orthorhombic phase and film ferroelectric polarization were recently demonstrated, but the impacts on endurance and retention are unknown. Remarkably, epitaxial HZO films on Si(001), where the stress caused by thermal mismatch likely favors epitaxial stabilization of the orthorhombic phase, present better endurance and retention than equivalent films on SrTiO$_3$(001).[11] On the other hand, the coercive field of epitaxial films is larger than that of equivalent polycrystalline films.[11–13,17,24–27] Understanding the reasons for the large $E_c$ in epitaxial films could allow its reduction, eventually resulting in an enhancement of endurance. Finally, endurance in epitaxial hafnia films is only reported for Pt/Hf$_{0.5}$Zr$_{0.5}$O$_2$/LSMO capacitors[11,24,26] and it is known that Pt electrodes induce oxygen vacancies and fatigue in conventional perovskite ferroelectrics.[29] The endurance of ferroelectric polycrystalline hafnia can depend severely on oxygen vacancies[60] and, in epitaxial HZO, replacement of the Pt electrode with a conducting oxide could allow for reduced fatigue. In summary, several strategies can be explored to understand the fatigue mechanisms in epitaxial films with the final aim of further improving performance.

## Experimental

The HZO film and buffer layers were grown by pulsed laser deposition (KrF excimer laser) in a single process. HZO/LSMO/LNO/CeO$_2$/YSZ heterostructures were deposited on Si(001) in a single process by pulsed laser deposition (KrF excimer laser). The Si(001) substrate was used as received. Substrate temperature (measured with a thermocouple inserted in the heater block) was 800 °C for YSZ and CeO$_2$ and 700 °C for LNO and LSMO. During deposition, there was a dynamic oxygen pressure of $4 \times 10^{-4}$ mbar for YSZ and CeO$_2$, 0.15 mbar for LNO, and 0.1 mbar for LSMO. The laser frequency for the deposition of buffer layers and LSMO electrode was 5 Hz. HZO films of varied thickness were prepared at $T_s = 800$ °C, PO$_2$ = 0.1 mbar, and laser frequency of 2 Hz, controlling the thickness (in the 4.6–18.4 nm range) by the number of laser pulses (from 400 to 1600). At the end of the deposition, samples were cooled under 0.2 mbar oxygen pressure. Platinum circular top electrodes, of diameter 20 µm and thickness 20 nm, were deposited by sputtering through stencil masks (Fig. 1a).

XRD was performed with Cu Kα radiation using a Siemens D5000 diffractometer equipped with a point detector and a Bruker D8-Advance diffractometer equipped with a 2D detector. Surface topography was studied using AFM.

Ferroelectric polarization loops were obtained at 1 kHz using the dynamic leakage current compensation (DLCC) procedure[61,62] at room temperature in top-bottom configuration by means of an AixACCT TFAnalyser2000 platform. Ferroelectric endurance was measured by cycling the sample at a frequency of 100 kHz using bipolar square pulses of indicated amplitude and measuring the polarization loop in the conditions indicated above, from which remanent polarization and imprint electric field were determined (see sketch in Fig. S13, ESI†). No variation of endurance results has been observed using a frequency lower than 100 kHz, indicating that the reported values are not influenced by experimental set-up or ferroelectric switching time response (Fig. S14, ESI†). Precisely at the used 100 kHz cycling frequency, an almost complete ferroelectric switching was observed during each bipolar square pulse (Fig. S15, ESI†) without any influence on endurance characterization. Retention time was measured by poling the sample using a triangular pulse of 0.25 ms and the indicated amplitude and determining the remanent polarization and imprint electric field from the first polarization curve of the polarization loop measured at 1 kHz using the selective PU and ND protocol[61] (see sketch in Fig. S16, ESI†) after the indicated time (see measurements of the $t = 9.7$ nm film in Fig. S17, ESI†). Capacitance ($C$) loops were measured using an impedance analyzer (HP4192LF, Agilent Co.) operated with an excitation voltage of 50 mV at 50 kHz. Relative dielectric permittivity ($\varepsilon_r$)-voltage loops were extracted from capacitance values using the $C = \varepsilon_0\varepsilon_r A/t$ relation, where $A$ is the electrode area and $t$ is the film thickness.

## Conclusions

In conclusion, a combination of high polarization, endurance and retention in epitaxial HZO films on Si(001) has been demonstrated. The properties depend critically on the HZO thickness and robust ferroelectricity is maintained down to the thinnest thickness investigated, below 5 nm. This achievement, using epitaxial films, proves that the ferroelectric properties of polycrystalline films of the same Hf$_{0.5}$Zr$_{0.5}$O$_2$ composition can be improved significantly. Our results signal that epitaxial films can help overcome the most serious weakness of polycrystalline doped HfO$_2$, namely the limited endurance and the polarization-endurance-retention dilemma.

## Conflicts of interest

There are no conflicts to declare.

## Acknowledgements

Financial support from the Spanish Ministerio de Ciencia e Innovación, through the "Severo Ochoa" Programme for Centres of Excellence in R&D (SEV-2015-0496) and the MAT2017-85232-R (AEI/FEDER, EU), and MAT2015-73839-JIN projects, and from Generalitat de Catalunya (2017 SGR 1377) is acknowledged. IF acknowledges Ramón y Cajal contract RYC-2017-22531. JL and TS are financially supported by China Scholarship Council (CSC) with No. 201506080019 and 201807000104, respectively. JL and TS work has been done as a part of their Ph.D. program in Materials Science at Universitat Autònoma de Barcelona. We acknowledge support of the publication fee by the CSIC Open Access Publication Support Initiative through its Unit of Information Resources for Research (URICI).

## Notes and references